\documentclass[twocolumn]{aastex62}

\graphicspath{{./}{figures/}}

\submitjournal{ApJL}

\shorttitle{low-temperature rate coefficients for CN + C$_6$H$_6$}
\shortauthors{Cooke et al.}

\begin{document}

\title{Benzonitrile as a proxy for benzene in the cold ISM: low temperature rate coefficients for CN + C$_6$H$_6$}

\correspondingauthor{Ian R. Sims}
\email{ian.sims@univ-rennes1.fr}

\author[0000-0002-0850-7426]{Ilsa R. Cooke}

\affil{Univ Rennes, CNRS, IPR (Institut de Physique de Rennes) - UMR 6251, F-35000 5 Rennes, France}

\author[0000-0002-6639-4909]{Divita Gupta}
\affil{Univ Rennes, CNRS, IPR (Institut de Physique de Rennes) - UMR 6251, F-35000 5 Rennes, France}

\author{Joseph P. Messinger}
\affil{Univ Rennes, CNRS, IPR (Institut de Physique de Rennes) - UMR 6251, F-35000 5 Rennes, France}
\affiliation{Division of Chemistry and Chemical Engineering, California Institute of Technology, Pasadena, California 91125, United States}

\author[0000-0001-7870-1585]{Ian R. Sims}
\affil{Univ Rennes, CNRS, IPR (Institut de Physique de Rennes) - UMR 6251, F-35000 5 Rennes, France}



\begin{abstract}

The low temperature reaction between CN and benzene (C$_6$H$_6$) is of significant interest in the astrochemical community due to the recent detection of benzonitrile, the first aromatic molecule identified in the interstellar medium (ISM) using radio astronomy. Benzonitrile is suggested to be a low temperature proxy for benzene, one of the simplest aromatic molecules, which may be a precursor to polycyclic aromatic hydrocarbons (PAHs). In order to assess the robustness of benzonitrile as a proxy for benzene, low temperature kinetics measurements are required to confirm whether the reaction remains rapid at the low gas temperatures found in cold dense clouds. Here, we study the C$_6$H$_6$ + CN reaction in the temperature range 15--295 K, using the well-established CRESU technique (a French acronym standing for Reaction Kinetics in Uniform Supersonic Flow) combined with Pulsed Laser Photolysis-Laser-Induced Fluorescence (PLP-LIF).
We obtain rate coefficients, $k(T)$, in the range (3.6--5.4) $\times$ 10$^{-10}$ cm$^3$ s$^{-1}$ with no obvious temperature dependence between 15--295 K, confirming that the CN + C$_6$H$_6$ reaction remains rapid at temperatures relevant to the cold ISM.
\end{abstract}

\keywords{Reaction rates (2081), Interstellar molecules (849), Astrochemistry (75), Interdisciplinary astronomy (804), Polycyclic aromatic hydrocarbons (1280), Dense interstellar clouds (371), Interstellar clouds (834), Interstellar medium (847)}


\section{Introduction} \label{sec:intro}

The recent discovery of benzonitrile (C$_6$H$_5$CN) in a nearby cold molecular cloud (Taurus, TMC-1) marks the first detection of an aromatic species in the interstellar medium by radio astronomy \citep{McGuire2018}. Benzonitrile has been proposed as a tracer of benzene, which may be a low-temperature precursor to more complex polycyclic aromatic hydrocarbons (PAHs). PAHs are widely accepted to exist in the ISM owing to their characteristic infrared emission features; however, due to their structural similarities, the exact chemical origins of the IR bands remain elusive \citep{Lovas2005}. It has been suggested that PAHs make up as much as 10--25\% of interstellar carbon budget \citep{Dwek1997,Chiar2013}. Understanding the origin of PAHs can help answer fundamental questions about their role in forming interstellar dust as well as potentially prebiotic material that may be incorporated into new planetary systems. 

It has been suggested that a bottleneck to the formation of these large aromatics at low temperatures is the cyclization to produce the first aromatic ring, usually benzene or the phenyl radical \citep{Cherchneff1992,Tielens1997,Kaiser2015}. However, benzene itself is difficult to detect as it has no permanent dipole moment and hence is invisible to radio astronomy. While benzene has been detected in the ISM through infrared observations of a single weak absorption feature (the $\nu_4$ bending mode near 14.85 mm) in a handful of sources \citep{Cernicharo2001,Kraemer2006,Malek2011}, absorption due to Earth's atmosphere limits its observation to bright IR-sources using space-based infrared telescopes (e.g. the Spitzer Space telescope and the Infrared Space Observatory).  Instead, it was suggested that benzonitrile (dipole moment = 4.5 D) might be used as a chemical proxy for benzene in the cold, starless ISM, as it is expected to form via the barrierless, exothermic neutral-neutral reaction between CN and benzene \citep{Woods2002,Trevitt2010}. If linked to benzene, benzonitrile may be used to constrain the early stages of the aromatic chemistry in the ISM. 

Following a tentative detection using data from the Nobeyama 45 m telescope, \citet{McGuire2018} conducted a deep integration search for benzonitrile in TMC-1 with the 100 m Robert C. Byrd Green Bank Telescope (GBT). They observed eight rotational transitions between 18 and 23 GHz, several of which displayed resolved hyperfine splitting, confirming the detection of benzonitrile and allowing derivation of its column density, $N_T$ = 4$\times$10$^{11}$ cm$^{–2}$.

\citet{McGuire2018} modified the KIDA 2014 reaction network \citep{Wakelam2015} to include the CN + C$_6$H$_6$ reaction, as well as a handful of destruction pathways in order to predict the abundance of benzonitrile in TMC-1. Currently, the KIDA database (kida.obs.u-bordeaux1.fr, accessed January 2020) does not include benzonitrile but does include 13 ion-neutral pathways for benzene destruction and only five neutral-neutral pathways  (mostly H-abstraction mechanisms). \citet{McGuire2018} used a reaction rate coefficient for CN + C$_6$H$_6$ of 3 $\times$ 10$^{-10}$ cm$^3$ s$^{-1}$, based on the assumption that the reaction will occur upon every collision. Their model predicted a benzonitrile abundance a factor of four lower than what was observed in TMC-1. The authors suggested that the  difference may be explained by other formation routes for benzene and/or benzonitrile that were not considered in their model.

Rate coefficients for the reaction between CN and benzene have been previously measured using PLP-LIF at 295, 165 and 105 K by \citet{Trevitt2010}, who found $k(T)$ to be relatively constant, ranging from 3.9--4.9 $\times$ 10$^{-10}$ cm$^{3}$ molecule$^{-1}$ s$^{-1}$. They attribute the lack of temperature dependence to a reaction mechanism that proceeds without an energy barrier in the entrance channel and likely forms an addition complex; which is supported by theoretical calculations of the reaction potential energy surface \citep{Balucani1999,Woon2006,Lee2019}.
In addition to the overall rate coefficients, \citet{Trevitt2010} measured the products of the reaction at room temperature using synchrotron VUV photoionization mass spectrometry. They found that the photoionization efficiency curve at $m/z$=103 could be well fit by benzonitrile photoionization spectrum with no detectable evidence for the H-abstraction product channel, C$_6$H$_5$ + HCN, nor the -NC isomer phenylisocyanide. 

Here, we report rate coefficients for the reaction of benzene with the CN radical over a wide temperature range of 15--295 K using the CRESU technique. These rate coefficients can be input into astrochemical models to assess the importance of the reaction as a production route for benzonitrile and therefore the robustness of benzonitrile as a chemical proxy for benzene at various temperatures in the ISM.

\section{Experimental}

\begin{figure*}[ht!]
    \centering
    \includegraphics[scale =0.3]{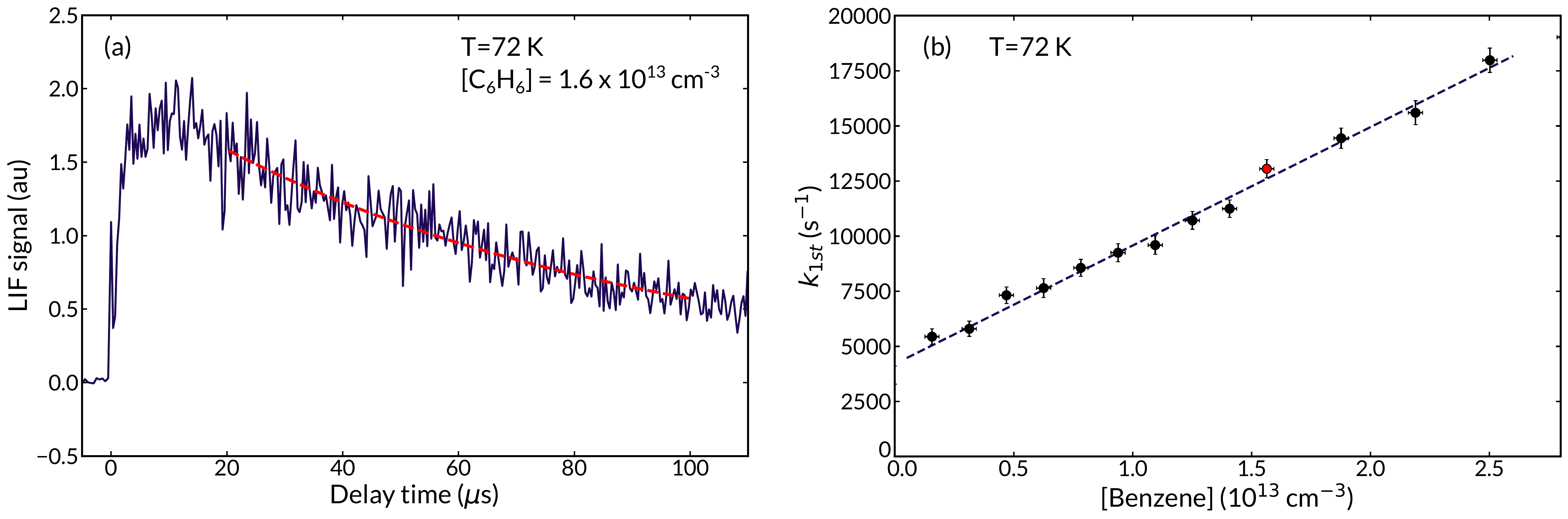}
	\caption{Typical kinetics data obtained using PLP-LIF in continuous CRESU flows, shown for a temperature of 72 K. \textit{(a)} LIF signal decay of CN($B^2\Sigma^+ \rightarrow X^2\Sigma^+$) at a benzene density of 1.6$\times$10$^{13}$ cm$^{-3}$ fit with an exponential decay function (red dashed line). \textit{(b)} second-order kinetics plot where the pseudo-first order rate coefficients are plot against the benzene concentration. The dashed line shows a linear least squares regression to the data yielding the second-order bimolecular rate coefficient. The red point corresponds to the k$_{1st}$ obtained from (a).}
	\label{fig:LIF}
\end{figure*}

The kinetics of the reaction between the CN radical and benzene were measured using Pulsed-Laser Photolysis-Laser Induced Fluorescence (PLP-LIF). The low gas temperatures were achieved using the CRESU technique (\textit{Cin\'{e}tique de Reaction en Ecoulement Supersonique Uniforme}; Reaction Kinetics in Uniform Supersonic Flow), which has been described in detail previously \citep{Sims1994,James1998,Cooke2019} and will only be described here in brief. A dilute mixture (typically $<$0.1\%) of benzene (Sigma Aldrich, Anhydrous 99.8\%) and the CN radical precursor in a buffer gas (He (99.995\%), Ar (99.998\%), or N$_2$ (99.995\%); Air Liquide) expands isentropically from a high-pressure region (reservoir), into a low-pressure region (chamber) through a convergent-divergent Laval nozzle to generate a cold supersonic flow that is uniform in temperature and density for several tens of centimeters, corresponding typically to 100--500 $\mu$s. The high molecular density of the flow (10$^{16}$--10$^{17}$ cm$^{-3}$) results in collisions between the molecules, ensuring thermal equilibrium is maintained. A wide range of temperatures can be obtained by careful manipulation of the physical dimensions of the nozzle, the buffer gas used, and the flow conditions/pumping capacity. Characterizing the flow temperature is essential to study chemical kinetics and is determined using Pitot probe impact pressure measurements for each nozzle prior to undertaking kinetics experiments.

A Controlled Evaporation and Mixing system was used to introduce benzene into the flow (Bronkhorst CEM, described in detail in \citet{Gupta2019}). The system consists of a Coriolis liquid flow meter (Bronkhorst mini CORI-FLOW ML120V00) that introduces a metered flow of liquid benzene into an evaporator (Bronkhorst W-202A-222-K), followed by controlled dilution with a buffer gas via a thermal mass flow controller (Bronkhorst EL-FLOW Prestige). The CN radicals were generated by photodissociation of ICN (Acros Organics, 98\%) at 248 nm using a KrF excimer laser (Coherent LPXPro 210) at a laser fluence in the reaction zone of $\sim$35 mJ cm$^{-2}$. Photodissociation of ICN at 248 nm produces CN radicals primarily ($\geq$93\%) in the v=0 level of the \textit{X$^{2}\Sigma^{+}$} state \citep{OHalloran1987} and rapid rotational relaxation is ensured by frequent collisions due to the high density of the buffer gas.

The LIF/fluorescence signal was recorded with a photomultiplier tube (PMT, Thorn EMI 6723) by excitation in the (0,0) band of the B$^{2}\Sigma^{+}$--X$^{2}\Sigma^{+}$ electronic transition at $\sim$388 nm using a dye laser (Laser Analytical Systems, LDL 20505, operating with 0.2g/L mixture of Exciton Exalite 389 dye in 1,4-dioxane) pumped by the frequency-tripled output at 355 nm of a Nd:YAG laser (Continuum, Powerlite Precision II). A narrowband interference filter centered at 420 nm (Ealing Optics, 10 nm FWHM) was used to select the off-resonant fluorescence into the first vibrational level of the ground state via the (0,1) band. Delays ranging from -5 to hundreds of microseconds between the excimer and the LIF dye laser pulses, both operating at 10 Hz, were employed to record the time dependence of the CN radical fluorescence during the reaction.

The experiments were performed under pseudo-first order conditions with benzene in excess. The UV absorption cross-section of ICN at 248 nm has been measured as 4.7 $\times$ 10$^{-19}$ cm$^2$ \citep{Felps1991}. For the density of ICN used in the gas flow ($\sim$10$^{12}$ cm$^{-3}$), there is an estimated CN (X$^{2}\Sigma^{+}$) concentration of $<$1$\times$10$^{10}$ cm$^{-3}$, which is lower than the concentration of benzene by at least a factor of one hundred.

The LIF signals were recorded using a gated integrator (Stanford Research Systems) at 400 evenly-spaced time delays and were averaged typically 5 times. These were fit to single exponential decays using Scipy's curve fit optimization package \citep{Virtanen2019}, yielding pseudo-first order rate coefficients, $k_{1st}$. Plots of $k_{1st}$ versus the benzene concentration were the fit with a weighted linear regression (Skipper \textit{et. al.} 2017, statsmodels, v0.10.2, Zenodo, 10.5281/zenodo.275519, as developed on GitHub) to calculate the second-order rate coefficient. The procedure was repeated for various temperatures using eight different Laval nozzles. The experimental parameters and the measured rate coefficients with their uncertainties are summarized in Table 1. 

\section{Results}

Figure \ref{fig:LIF}(a) shows an example CN LIF decay trace and second order kinetics plot, obtained at 72 K. The signal decays were recorded for 140--200 $\mu$s following the firing of the photolysis laser to capture the fast CN decay in the presence of benzene.  LIF measurements were also taken at negative time delays (5 $\mu$s before the excimer laser fires) to establish a pre-trigger baseline. An exponential decay function was fit to the LIF data after $\geq$10 $\mu$s to allow enough time for rotational relaxation of CN and for the photomultiplier tube to recover from scatter due to the excimer laser. 

The LIF decay traces were recorded for the CN decay in the presence of at least eight different concentrations of benzene. The number of points and the range of benzene concentration used for each nozzle are shown in Table 1. The upper limit for the concentration of benzene is determined by an experimental constraint imposed due to the formation of benzene dimers, which could cause the rate coefficients to be underestimated. This effect is particularly significant at low temperatures, where dimerization occurs at much lower reactant concentrations than at room temperature.  \citet{Hamon2000} measured the rate coefficients for dimerization of benzene in helium. They found that the onset of significant complex formation at 25 K occurred when [C$_6$H$_6$] $\geq$1 $\times$ 10$^{14}$ molecule cm$^{-3}$. We obtain the bimolecular rate coefficients by fitting benzene concentrations $<$2.5 $\times$ 10$^{13}$ molecules cm$^{-3}$, except for at room temperature, where the fit concentrations are $<$1 $\times$ 10$^{14}$ molecules cm$^{-3}$.

\begin{figure*}[t!]
    \centering
    \includegraphics[scale =0.35]{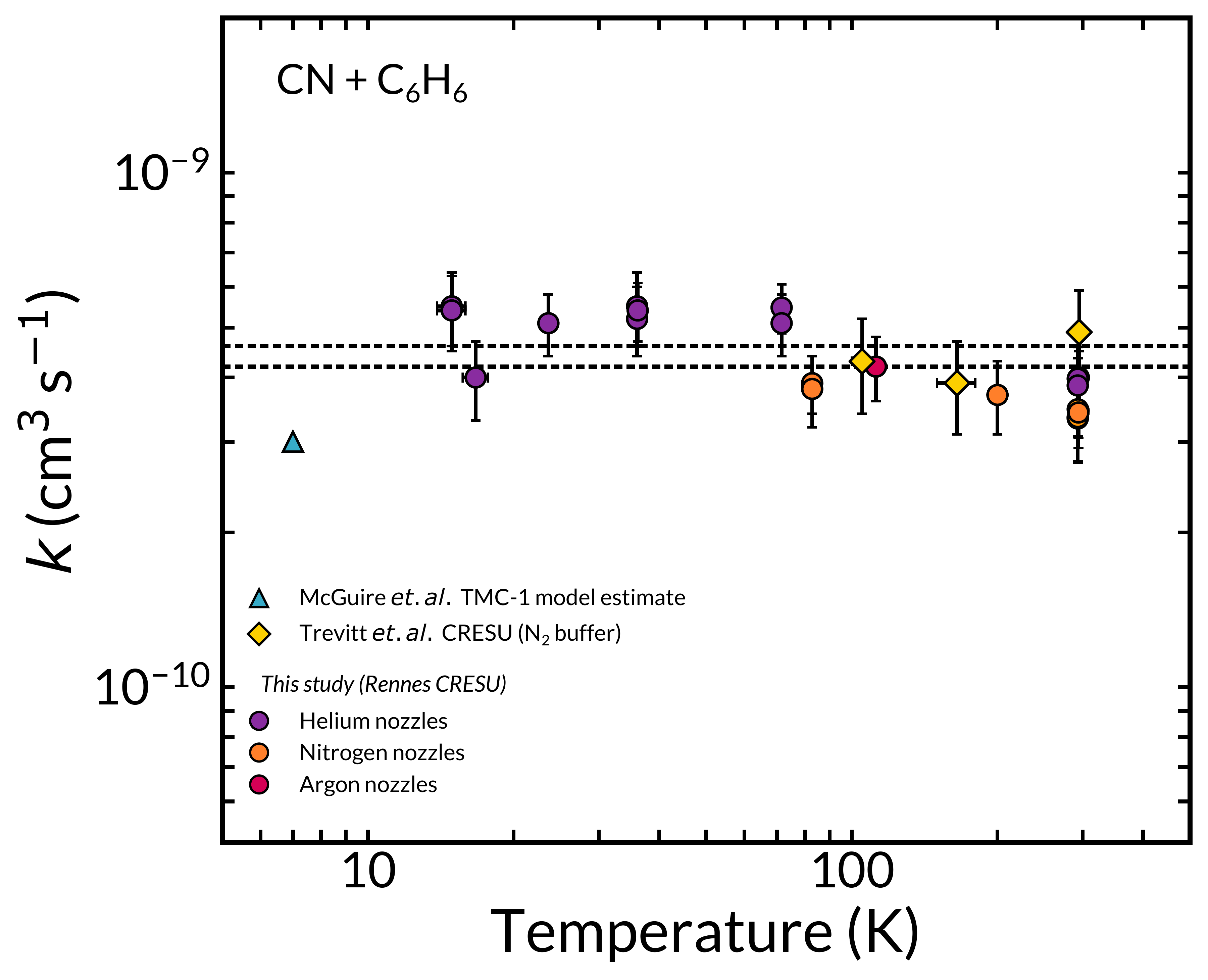}
	\caption{Rate coefficients for the reaction of the CN radical with benzene as a function of temperature displayed on a log-log scale. The circles are the data taken here using the continuous CRESU in Rennes, with helium buffer gas in purple, nitrogen in orange and argon in pink. The yellow diamonds are the data taken by \citet{Trevitt2010} using a pulsed CRESU. The teal triangle shows the estimated rate coefficient used by \citet{McGuire2018} to model the abundance of benzonitrile in TMC-1 at 7 K. The dashed lines show the upper and lower limits set by the standard error ($\pm$1$\sigma$) on the variance weighted mean of the readings at all temperatures.}
	\label{fig:Tdep}
\end{figure*}

The non-zero intercept observed in the second order plots (e.g. in Figure \ref{fig:LIF}(b)) is due to other losses of CN, a combination of CN diffusion out of the probed beam area and reaction with ICN and/or other impurities in the buffer gas. The second-order rate coefficients are derived from the slopes of the weighted linear-least squares regressions to the 2nd-order kinetics plots and are shown in Table 1 along with their uncertainties, which are calculated using 95\% confidence limits of the standard error from the two-sided Student's t-distribution combined with an estimated 10\% systematic error. The measured rate coefficients over the temperature range $T$ = 15--295 K are shown in Figure \ref{fig:Tdep} along with the rate coefficients measured by \citet{Trevitt2010} and the rate coefficient used by \citet{McGuire2018} to model the abundance of benzonitrile in TMC-1. Within the experimental uncertainty the measured rate coefficient for the CN + C$_6$H$_6$ reaction remains essentially constant between 15--295 K. The weighted average of all of the rate coefficients between 15--295 K is 4.4 $\pm$ 0.2 $\times$ 10$^{-10}$ cm$^3$ s$^{-1}$.


\begin{deluxetable*}{cccccc}
    \tablecaption{Rate coefficients for the reaction of the CN radical with benzene measured at different temperatures, with the associated experimental parameters. Quoted uncertainties are calculated using the standard error evaluated from the second order plot, multiplied by the appropriate Student's t factor for 95\% confidence, and then combined in quadrature with an estimated systematic error of 10\%. Entries in bold are the variance weighted mean values of rate coefficients measured at the same temperature, where the standard error on the mean is combined in quadrature with an estimated systematic error of 10\%. \label{tab:data}}
	\tablecolumns{6}
	\tablewidth{0pt}
	\tablehead{
	    \colhead{T } &
	    \colhead{Buffer gas} &
	    \colhead{Total density} &
	    \colhead{Range of [C$_6$H$_6$]} &
	    \colhead{No. of points} &
	    \colhead{Rate coefficient, $k(T)$}\\
	    \colhead{(K)} & \colhead{} & \colhead{(10$^{16}$ cm$^{-3}$)} & \colhead{(10$^{12}$ cm$^{-3}$)} & \colhead{} & \colhead{(10$^{-10}$ cm$^3$ s$^{-1}$)}
	}
	\startdata	
	15 & He & 5.02 & 2.19--19.7 & 13 & 5.45 $\pm$ 0.90\\
	     &    &      & 2.19--17.5 & 14 & 5.36 $\pm$ 0.86 \\
	     &    &      &            &     & \textbf{5.4 $\pm$ 0.6} \\
	\hline
	17 & He & 4.85 & 2.03--14.2  & 13 & 4.5 $\pm$ 0.7\\
	\hline
	24 & He & 4.85 & 2.08--22.8 & 11 & 5.1 $\pm$ 0.7\\
	\hline
	36 & He & 5.31 & 1.48--14.8 & 11 & 5.37 $\pm$ 0.70\\
	   &    & 5.27 & 1.47--20.6 & 14 & 5.49 $\pm$ 0.94 \\
	   &    & 5.27 & 1.47--14.7 & 11 & 5.18 $\pm$ 0.77\\
	   &    &      &            &    & \textbf{5.3 $\pm$ 0.6} \\
	\hline
	72 & He & 6.01 & 1.54--24.6 & 14 & 5.47 $\pm$ 0.63\\
	   &  & 6.01 & 3.00--16.5 & 12 & 5.12 $\pm$ 0.73 \\
	   &    &      &            &    & \textbf{5.4 $\pm$ 0.6} \\
	\hline
	83 & N$_2$ & 4.61 & 2.06--16.5 & 8 & 3.89 $\pm$ 0.70\\
	     &       &      & 2.04--24.6 & 12 & 3.85 $\pm$ 0.62 \\
	     &       &      &            &    & \textbf{3.9 $\pm$ 0.4}\\
	\hline
	110  & Ar    & 2.71 & 1.66--19.9 & 11 & 4.2 $\pm$ 0.6 \\
	\hline
	200 & N$_2$  & 5.27 & 2.52--20.2 & 13 & 3.7 $\pm$ 0.8 \\
	\hline
	294 & N$_2$  & 7.04 & 1.52--76.1  & 9 & 3.47 $\pm$ 0.40 \\
	293 & N$_2$  & 9.78 & 1.53--76.6  & 9 & 3.45 $\pm$ 0.39 \\
	    & N$_2$  & 18.5 & 2.21--69.6  & 8 & 3.33 $\pm$ 0.54\\
	293 & He     & 5.27 & 6.25--37.5  & 9 & 3.98 $\pm$ 0.57\\
	    & He     & 6.92 & 2.03--81.3  & 10 & 3.35 $\pm$ 0.46\\
	    & He     & 9.39 &  2.16--63.4 & 6 & 4.08 $\pm$ 0.55 \\
	295 & He    & 9.10 &  1.50--74.8 &  9   & 4.02 $\pm$ 0.46\\
	   &        &      &             &      & \textbf{3.6 $\pm$ 0.4}\\
	\enddata
\end{deluxetable*}

Care was taken to ensure photodissociation of C$_6$H$_6$ was negligible during the kinetics experiments as reaction between the phenyl radical and CN is expected to be rapid. The absorption cross section for benzene at 248 nm is 1.4 $\times$ 10$^{-19}$ cm$^2$ \citep{Nakashima1982} yielding a maximum phenyl radical concentration of less than 10$^{10}$ cm$^{-3}$ at the laser fluence used of $\sim$25 mJ/cm$^2$. \citet{Kovacs2009} have suggested the possibility of two-photon absorption via the $^1B_{2u}$ state, leading to dissociation of benzene; they found a high total absorption cross section for the second photon of 2.8 $\times$ 10$^{-17}$ cm$^2$. For the highest benzene concentration used here ($\sim$1 $\times$ 10$^{14}$ cm$^{-3}$) and with the laser fluence used, all photolysis of benzene occurs by two-photon dissociation. The resulting phenyl radical concentration is too low ($<$1\%) to affect the kinetics of the reaction. The concentrations of the phenyl radical are even lower in reality due to quenching of the excited $^1B_{2u}$ state by the buffer gas. To experimentally verify this, we measured the first-order rate coefficient at $T$ = 22.9 K and [C$_6$H$_6$] = 2$ \times$ 10$^{13}$ cm$^{-3}$ while varying the excimer laser fluence over the range of 19--30 mJ/cm$^2$ and found that $k_{1st}$ remained constant.

\section{Discussion}

\subsection{Chemical kinetics}

The rate coefficients for the reaction of CN with C$_6$H$_6$, measured here between 15 and 295 K, are consistent with those previously measured by \citet{Trevitt2010} at 295, 165 and 105 K with a pulsed CRESU apparatus and PLP-LIF. \citet{Trevitt2010} predicted the reaction would remain rapid at even lower temperatures, as confirmed by our measurements. 
The lack of temperature dependence observed in both studies is consistent with a barrierless entrance channel and the formation of an addition complex. It has been shown both experimentally and theoretically that addition-elimination reactions between CN and unsaturated hydrocarbons typically have no barriers in their entrance channel due to the formation of intermediate radical adducts that are relatively stable \citep{Carty2001}. In general, for this class of reactions, there is usually at least one exothermic exit channel to products. \citet{Goulay2006} found a similarly large rate coefficient for the reaction between the ethynyl radical (C$_2$H) and benzene with essentially no temperature dependence between 105--298 K.

The rate coefficients presented here and those of \citet{Trevitt2010} are similar to those of \citet{Woon2006}, who used trajectory calculations to predict the rate coefficients between 50--300 K. The calculated rate coefficients ranged from 3.15--3.5 $\times$ 10$^{-10}$ cm$^{3}$ molecule$^{-1}$ s$^{-1}$ when back reactions to reform CN + C$_6$H$_6$ are excluded. \citet{Woon2006} also investigated the pressure dependence of the reaction by including back-dissociation of the C$_6$H$_6$-CN intermediate complex using a multiwell treatment and found a slight temperature dependence at 10$^{-3}$ mbar, which converges with the high pressure limit rate coefficient by 50 K. We measured the room temperature rate coefficient at three different pressures for both nitrogen and helium buffer gases and did not observe a pressure dependence. It is important to note, however, that the laboratory experiments were likely conducted within the high pressure limit.  

\subsection{Reaction products}

Previous theoretical and experimental studies have shown that benzonitrile is the major species produced in the reaction between CN and benzene, with negligible or no production of the -NC isomer phenylisocyanide nor the H-abstraction product, the phenyl radical. 
\citet{Woon2006} calculated the reaction potential energy surface and relative product yields and found that while the C$_6$H$_5$CN + H products form exothermically and barrierlessly, there is a 25 kJ/mol barrier to form the isocyano product pair C$_6$H$_5$NC + H after initial barrierless formation of the C$_6$H$_5$NC adduct, which was found to be extremely rare in multiwell calculations. 

Crossed beam experiments conducted under single collision conditions at much higher energies have demonstrated that the benzonitrile is the main reaction product \citep{Balucani1999,Balucani2000,Balucani2000a}. \citet{Balucani1999} conducted crossed beam experiments at collision energies between 19.5 and 34.4 kJ mol$^{-1}$ as well as electronic structure and RRKM calculations. Neither the C$_6$H$_6$CN adduct nor the phenylisocyanide isomer C$_6$H$_5$NC were found to contribute to the crossed beam scattering signal. It was concluded that the dominant reaction entrance channel is barrierless leading to the formation of a CN-addition complex that subsequently dissociates to form benzonitrile + H, with the C$_6$H$_5$NC product channel contributing less than 2\%. 

\citet{Lee2019} studied the reaction of benzene + CN indirectly using microwave discharge experiments and cavity Fourier transform microwave spectroscopy, in combination with electronic structure calculations. They found that the reaction produces benzonitrile in high yield, with $<$0.1\% relative abundance of phenylisocyanide. Isotopic measurements confirmed that the CN bond remains intact during the product formation. In contrast to previous computations, they found formation of the \textit{iso-}adduct requires surmounting a barrier of $\sim$17 kJ/mol as well as a second barrier for H atom loss; implying that phenylisocynaide formation should be highly disfavored under low temperature conditions. While these very sensitive experiments provide important insights into possible reaction pathways, it is important to note that they were not conducted under conditions of kinetic isolation, nor at thermal equilibrium. Under these conditions it is difficult to probe any specific reaction mechanism as multiple reaction paths may contribute and it is unclear how well the product branching ratios obtained reflect those that would be obtained for the elementary reaction CN + C$_6$H$_6$ at a well-defined temperature. Low-temperature kinetics experiments are still needed to identify and confirm the product channel specific rate coefficients that are critical parameters for models of astrochemical environments and planetary atmospheres. 

\begin{figure}[t!]
    \centering
    \plotone{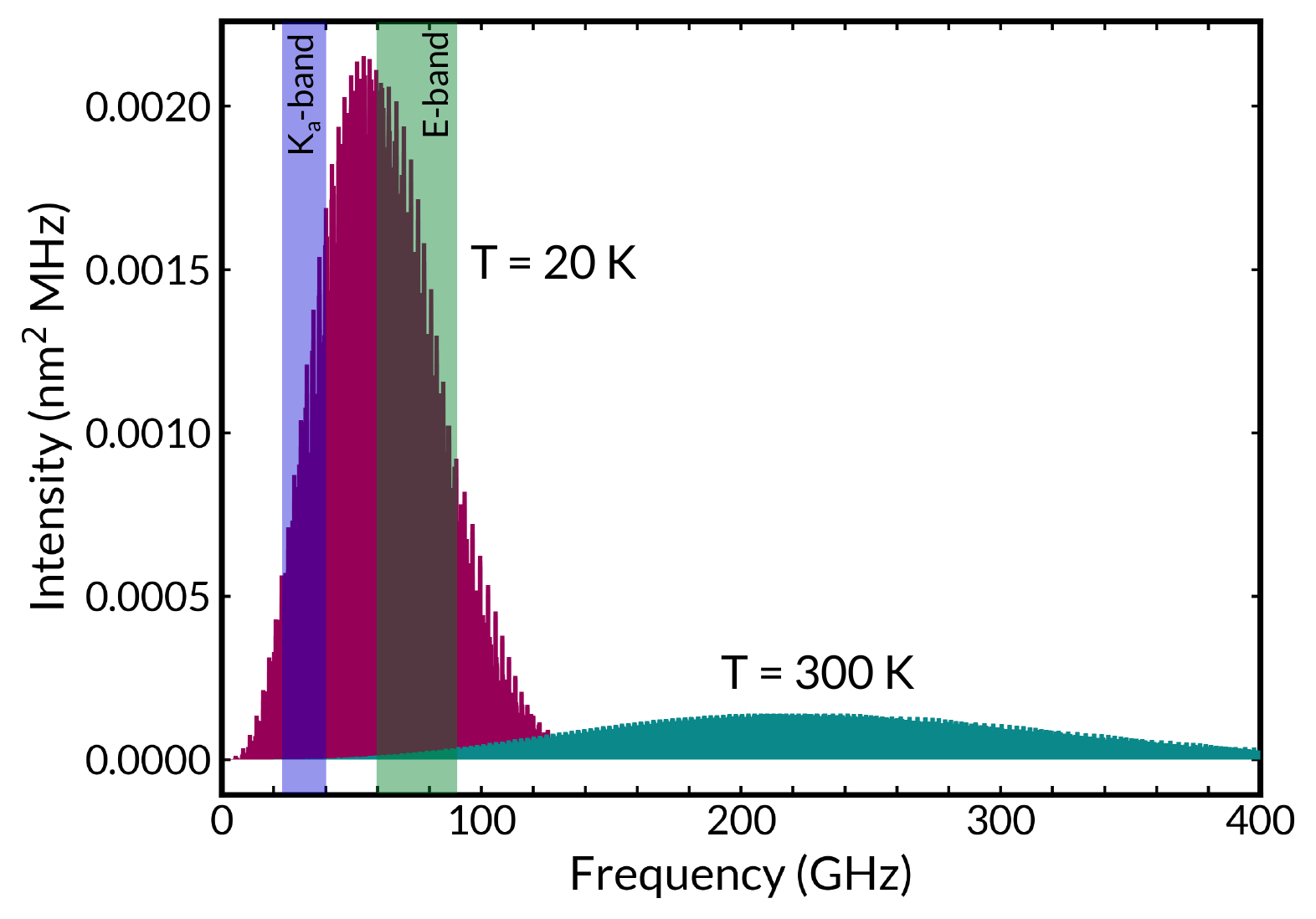}
	\caption{Simulated benzonitrile spectra at 20 K (magenta) and room temperature (teal). The vertical blocks show the bandwidth of the two CPUF spectrometers in Rennes, one covering the Ka band from 26.5--40 GHz and the other the E-band from 60--90 GHz. The spectra were simulated using data from the Cologne Database for Molecular Spectroscopy and Pickett's spfit/spcat program \citep{Pickett1991}.}
	\label{fig:spectra}
\end{figure}

Chirped-Pulse Fourier-Transform Microwave (CP-FTMW) spectroscopy has recently been combined with CRESU flows to measure the channel-specific rate coefficients at low temperature \citep{Abeysekera2015}. This technique (named Chirped-Pulse in Uniform flows, or CPUF) can be used to establish product-branching ratios at low temperatures, which are critical for astrochemical models but are challenging to obtain experimentally. K$_a$-band ($\sim$28-40 GHz) and E-band ($\sim$60-90 GHz) spectrometers have been coupled to the continuous CRESU flows in Rennes that will allow the formation of benzonitrile in the CN + C$_6$H$_6$ reaction to be measured down to low temperatures. Figure \ref{fig:spectra} shows how the CRESU flow acts as an efficient rotational refrigerator, shifting the Boltzmann distribution of the rotational lines into the range of the chirped-pulse spectrometers in Rennes. While benzonitrile is expected to be the sole product of CN + C$_6$H$_6$, the CPUF technique can be used to measure the product branching ratios for other multi-channel reactions involving CN and aromatics.

\subsection{Astrophysical Implications}

The rate coefficients measured here for the CN reaction with benzene are consistent with (albeit somewhat higher than) the value used in the astrochemical model of \citet{McGuire2018} to predict the abundance of benzonitrile in TMC-1, $k(T)$ = 3$\times$10$^{-10}$ cm$^{3}$ molecule$^{-1}$ s$^{-1}$ (Shingledecker, private communication 2019). It is therefore unlikely that this reaction is the cause of the discrepancy between the abundance of benzonitrile in their model versus that observed in TMC-1. The large rate coefficients observed over the full temperature range suggests that the reaction should be rapid in astronomical sources that have sufficient density of benzene and CN. Instead, the discrepancy between the observed and modeled abundances is likely due to the underproduction of benzene; due to missing production routes and/or underestimation of the rates for those already in the reaction network.

A photochemical model has been recently developed by \citet{Loison2019} to investigate the production of aromatics in the atmosphere of Titan. While the model predicted significant formation of toluene and ethylbenzene, benzonitrile was not predicted to be abundant due to efficient consumption of CN by methane.
 
It is likely that other radicals can add to benzene with similarly rapid rate coefficients, indicating that other benzene derivatives may be present in the cold ISM. The reaction of CH with benzene was measured by \citet{Hamon2000} at 25 K and found to be similarly rapid with a rate coefficient of 2.7 $\times$ 10$^{-10}$ cm$^3$ s$^{-1}$. Rate coefficients for the reaction of OH with benzene down to 58 K were presented in a review by \citet{Hansmann2007} and were found to be an order of magnitude lower; however, a complete kinetics study has yet to be published. The reaction of CN with other aromatics may also be rapid and produce products with high dipole moments that could lend themselves to astronomical detection. An alternative route to benzonitrile formation could involve the radiative association reaction between the phenyl (C$_6$H$_5$) and CN radicals, which may be rapid under ISM conditions but has not yet been investigated. The reactions of CH and OH with anthracene have both been measured in CRESU flows \citep{Goulay2005,Goulay2006a}, though reactions of CN with aromatics larger than benzene have not yet been studied. 

The formation of benzene under interstellar conditions is not as well understood and is a major source of uncertainty in astrochemical models of aromatic chemistry. \citet{Jones2011} investigated benzene formation via the reaction of C$_2$H with 1,3-butadiene. Benzene was observed at significant fractions (30\%$\pm$10\%) with the dominant reaction product the thermodynamically less stable isomer, hexa-1,3-dien-5-yne. However, \citet{Lockyear2015} also studied the products of this reaction using synchrotron photoionization mass spectrometry. The photoionization spectra indicated that fulvene is the major reaction product, with a branching fraction of $\sim$60\%. They did not detect benzene as a product and placed an upper limit on the production of benzene and hexa-1,3-dien-5-yne isomers of 45\%.  \citet{Lee2019} likewise observed evidence for fulvene formation in a microwave discharge containing HC$_3$N (a C$_2$H precursor) and 1,3-butadiene. The discrepancy may be due the high collisional energy of the crossed-beam experiment, emphasizing the importance of studying reactions under both single collision and thermal conditions. The main formation route for benzene in the kida2014 network is the dissociative recombination of C$_6$H$_7^+$ with an electron. \citet{McGuire2018} also added the reaction between C$_2$H + 1,3-butadiene to the kida2014 network with a rate coefficient of 3 $\times$ 10$^{-10}$ cm$^3$ s$^{-1}$. It is therefore possible that the benzene abundance in their model is overestimated since they do not account for the low (or zero) fraction of benzene likely produced by this reaction at low temperatures. In addition, 1,3-butadiene has not been detected in the ISM and thus it is unknown whether the reaction could produce significant quantities of benzene in TMC-1.

\section{Conclusions}

The CRESU technique, combined with PLP-LIF, has been used to measure the kinetics of the reaction between CN and benzene over the temperature range 15--295 K. We find that the rate coefficients for this reaction do not display an obvious temperature dependence over this temperature range, confirming that the CN + C$_6$H$_6$ reaction will remain rapid at temperatures relevant to the cold ISM. These results suggest that benzonitrile is indeed a robust chemical proxy that can be used to infer the abundance of benzene from observations made using radio astronomy. They also indicate that the discrepancy between the observed and modeled abundances of benzonitrile in TMC-1 is likely due to missing or inaccurate kinetic data for benzene production routes.

\section{Acknowledgements}

The authors thank Jonathan Courbe, Jonathan Thi\'{e}vin, Didier Biet, Ewen Gallou and Alexandre Dapp for technical support. The authors thank Brett McGuire, Christopher Shingledecker and Mitchio Okumura for helpful discussions. JPM was supported by the National Science Foundation Graduate Research Fellowship (NSF GRFP) and the National Science Foundation Graduate Research Opportunities Worldwide (NSF GROW) programs. JPM would also like to thank the Office for Science and Technology of the Embassy of France in the United States for a Chateaubriand Fellowship. The authors acknowledge funding from the European Research Council (ERC) under the European Union’s Horizon 2020 research and innovation programme under grant agreement 695724-CRESUCHIRP and under the Marie Sklodowska-Curie grant agreement 845165-MIRAGE. The authors are also grateful for support from the European Regional Development Fund, the Region of Brittany and Rennes Metropole. This work was supported by the French National Programme ``Physique et Chimie du Milieu Interstellaire'' (PCMI) of CNRS/INSU with INC/INP co-funded by CEA and CNES.


\software{{\fontfamily{qcr}\selectfont SciPy} \citep{Virtanen2019}, {\fontfamily{qcr}\selectfont NumPy} \citep{VanDerWalt2011},  {\fontfamily{qcr}\selectfont Matplotlib} \citep{Hunter2007}, {\fontfamily{qcr}\selectfont StatsModels} (www.statsmodels.org)}

\end{document}